\documentclass[
  journal=pasa,
  manuscript=article-type,
  year=2020,
  volume=37,
]{cup-journal}

\usepackage{amsmath}
\usepackage[nopatch]{microtype}
\usepackage{booktabs}
\usepackage{mathtools}
\usepackage{amssymb}
\usepackage{tabularx}
\usepackage{multirow}
\usepackage{hyperref}
\usepackage{graphicx}
\usepackage{subcaption}
\usepackage{multirow}
\usepackage{tabularx}
\usepackage{booktabs}
\usepackage{caption}

\DeclareMathAlphabet\mathbfcal{OMS}{cmsy}{b}{n}

\title{Characterising and mitigating Bluetooth and WiFi radio frequency interference at the Parkes Observatory}

\author{Tommy Marshman}
\affiliation{School of Mathematical and Physical Sciences, and Astrophysics and Space Technologies Research Centre, Macquarie University, NSW 2109, Australia}
\alsoaffiliation{CSIRO Space and Astronomy, Australia Telescope National Facility, PO Box 76, Epping, NSW 1710, Australia}
\email[T. Marshman]{tommy.marshman@hdr.mq.edu.au}

\author{George Hobbs}
\affiliation{CSIRO Space and Astronomy, Australia Telescope National Facility, PO Box 76, Epping, NSW 1710, Australia}

\author{J. R. Dawson}
\affiliation{School of Mathematical and Physical Sciences, and Astrophysics and Space Technologies Research Centre, Macquarie University, NSW 2109, Australia}
\alsoaffiliation{CSIRO Space and Astronomy, Australia Telescope National Facility, PO Box 76, Epping, NSW 1710, Australia}

\author{Stefan  Osłowski}
\affiliation{CSIRO Space and Astronomy, Australia Telescope National Facility, PO Box 76, Epping, NSW 1710, Australia}

\author{John Tuthill}
\affiliation{CSIRO Space and Astronomy, Australia Telescope National Facility, PO Box 76, Epping, NSW 1710, Australia}

\author{Samantha Gordon}
\affiliation{CSIRO Space and Astronomy, Australia Telescope National Facility, PO Box 76, Epping, NSW 1710, Australia}

\author{John E. Reynolds}
\affiliation{CSIRO Space and Astronomy, Australia Telescope National Facility, PO Box 76, Epping, NSW 1710, Australia}

\author{Alex Dunning}
\affiliation{CSIRO Space and Astronomy, Australia Telescope National Facility, PO Box 76, Epping, NSW 1710, Australia}

 %% First letter not capped

\begin{document}

\begin{abstract}
We present a detailed characterisation of radio frequency interference (RFI) in the 2.4 GHz band around Murriyang, CSIRO’s Parkes radio telescope. The dominant sources of interference are Wi-Fi and Bluetooth transmissions. We quantify how the intensity and directionality of this RFI vary with time of day and document its evolution over several years. Although most observers currently discard data within this band, our analysis shows that the interference is confined in both time and frequency and can be effectively mitigated. Using 10 seconds of 16-bit voltage data recorded during observations of the Vela Pulsar (PSR~J0835$-$4510), we demonstrate that the majority of the channelised data remain unaffected by RFI. We compare three RFI detection and mitigation algorithms and evaluate their relative performance. All methods perform effectively, and any could be implemented in real time to enable productive use of this observing band. A real time implementation would allow the scientific use of this 128\,MHz observing band to increase, from almost 70\% of the band being completely unusable all of the time, to over 90\% of becoming accessible for science.  Given its simplicity and efficiency, a basic power-threshold approach offers a relatively straightforward solution.

\end{abstract}

\noindent 

\section{Introduction}
Radio frequency interference (RFI) continues to pose a significant challenge for radio astronomy. The prevalence and intensity of interference are increasing as populations expand near observatory sites, radio-emitting devices become more widespread, and the number of aircraft and satellites continues to grow. At the same time, radio astronomy is moving toward observations with increasingly wide bandwidths, further heightening its vulnerability to interference. Single-dish radio telescopes are particularly susceptible to these effects \citep{finger2017fpga,fridman2001rfi}. 

The majority of observations with the CSIRO Parkes 64-m radio telescope (also known as Murriyang in the local Wiradjuri language) are currently carried out using an ultra-wide-bandwidth low-frequency (UWL) observing system covering a frequency range between 704 to 4032\,MHz.  Details of the receiver and signal processing systems are provided in \citet{hobbs2020ultra}. Only a very small portion of this observing band is protected for radio astronomy. The band as a whole contains multiple forms of non-astronomical radio transmissions, from terrestrial transmitters as well as satellites and aircraft. There is also the potential for RFI to be generated from within the telescope system itself. 

Strong RFI can saturate the entire receiving system making the scientific observations unusable.  Weaker RFI can saturate parts of the signal path (e.g., a digitiser system).  All RFI within the observing band will reduce the science productivity of the telescope; the RFI can obscure the astrophysical signals or mimick possible astrophysical signals. `Perytons' were initially thought to be of extra-terrestrial origin due to their frequency-dispersed emission that mimicked the appearance of a broad-band signal that had traveled through cold plasma \citep{petroff2015identifying}. The actual source of Peryton emission was discovered to be the emission from the on-site microwave magnetron shutdown when the microwave door was prematurely opened during operation \citep{petroff2015identifying}. The Breakthrough Listen Candidate 1 (BLC1) \citep{sheikh2021analysis, smith2021radio} was detected during a Search for Extra-Terrestrial Intelligence (SETI) survey and was first thought to be a possible technosignature from an alien civilisation, but is likely to be locally produced RFI.

Ever-increasing levels of RFI have led to renewed interest in RFI mitigation techniques, so that the affected frequency ranges can once again be used for science. The UWL observing system was designed to ensure that, in most cases, neither the front-end receiver system, nor the backend digitiser system, would saturate in the presence of RFI.  The mitigation methods considered are therefore focused on removing the RFI from the observing band, whilst leaving any astronomical signal untouched.

In this work, we focus on the 2.4 GHz sub-band of the UWL, which spans 128 MHz from 2368 to 2496 MHz. This frequency range is heavily utilised by Wi-Fi and Bluetooth communications and also includes transmission signals from Australia’s National Broadband Network (NBN), the national open-access internet service. These are all major problems at Parkes, and only the 27\% of this subband between the WiFi/Bluetooth and NBN bands \citep{hobbs2020ultra} is normally used, with the remainder flagged as unusable for astronomical observations. The 2432\,MHz subband was chosen as it is the UWL subband that has the most data excised from observation, thus making a good testing ground for mitigation techniques. Potential astronomy applications in this band include: pulsar astronomy (since pulsars emit broadband signals), SETI \citep{worden2017breakthrough, isaacson2017breakthrough, price2018breakthrough}, and searching for Fast Radio Bursts (FRBs), which have been discovered around this range \citep{niu2021repeating}. 

The International Telecommunications Union  has designated the 2.4--2.5 GHz sub-band for use in industrial, scientific, and medical (ISM) applications \citep{ITU2020Radioregulations}. While the Australian Communications and Media Authority has added radio astronomy to the designation of this sub-band \citep{ACMA2021Spectrumplan} it is still heavily affected by RFI.  As wireless technology increases, there is an increasing demand for the commercial use of this part of the spectrum. Bluetooth operates in the 2.4\,GHz ISM band from 2.400 – 2.4835\,GHz \citep{Anees2024bluetooth}, while WiFi operates in a number of bands one of which is also the 2.4\,GHz ISM band. 

In this paper we aim to both characterise and mitigate the 2.4\,GHz Bluetooth and WiFi signals seen at the Parkes observatory site. There are many methods that have been proposed for RFI mitigation in radio astronomy, including different analogue and digital methods in both frequency and temporal domains, for a review of some of these methods see \cite{fridman2001rfi} and \cite{baan2019implementing}. As \cite{fridman2001rfi} note, there is no single method of RFI mitigation widely used in radio astronomy because the applicability and success of different methods will depend on the type of telescope (single dish or interferometer), the type of observations (e.g. folding at the pulsar period, spectral line, etc.), and the local RFI environment. 

For this paper we have obtained new observations, but also made use of archival observations.  These observations are described in \S2.  We use these observations to characterise the RFI in \S3. In \S4 we inspect the data sets with high frequency and time resolution and compare with the expected signal types from Bluetooth and WiFi. In \S5 we describe and compare different mitigation methods. We conclude in \S6.

\section{Observations}
\label{sec_obs}
For this paper, we make use of two types of data. The first are archival calibration observations that were primarily taken for pulsar observing projects. These files are folded observations of a pulsed noise source, which were taken in order to calibrate subsequent pulsar observations. These range from a few very short (few second duration) observations to one observed for almost 20 minutes. However, by far the majority had a duration of 2 minutes. These data files are in \textsc{psrfits} \citep{hotan2004psrchive} fold-mode data format, typically have 1\,MHz frequency resolution and are integrated over a few minutes.  We have chosen to make use of these files as (1) they are relatively small in volume and hence quick and easy to process and (2) provide almost daily snap-shots of the RFI environment over the entire band since the early observations with the receiver.  We process these data using the \textsc{pfits} and \textsc{psrchive} software suites.  Specifically we time- and polarisation- average using \textsc{pam} and then process the \textsc{dat\_offs} and \textsc{dat\_scls} \textsc{psrfits} tables in \textsc{pfits\_formBandpass} to determine the averaged spectral bandpass throughout the calibration observation. 

We record the meta-data (e.g., observation time and pointing direction) along with each spectrum and hence can determine the average signal strength in part of the band as a function of date, time-of-day, telescope Azimuth and Zenith, equatorial coordinates etc.  While we do not calibrate into physical units, we do account for known large-scale attenuation changes caused by the addition (or subtraction) of attenuator physical components into the signal chain.  All these observations are publicly available from the CSIRO Data Archive\footnote{\url{https://data.csiro.au}}.

It is not possible to use the calibration observations described above to test and demonstrate mitigation algorithms because they have been folded at the period of the switching noise source and channelised using a 2048-pt transform before summing the resulting powers of the two polarisation states into 1 MHz channels. We therefore have also recorded a 16-bit voltage data stream of our UWL subband of interest. During this 10\,s observation, taken on 10 August 2023 at UTC~02:41:26, the telescope was observing the bright pulsar PSR~J0835$-$4510 (also known as the Vela pulsar) and the corresponding data file was 10\,GB in volume. Vela was chosen as it is the brightest known pulsar and we can detect individual pulses from this pulsar in our 10\,s test data set.  As most of our mitigation methods are based on thresholding we can use this observation to demonstrate that we can remove bright RFI, without removing the bright astronomical signal (and noting that other pulsars will be weaker and hence even less affected by thresholding rejection methods). The corresponding data file is stored in the \textsc{dada} format as offset binary encoding, 2 polarisation states interleaved in blocks of 2048 voltage samples.  Such data can be processed using the \textsc{dspsr} software \citep{van2011dspsr}, but we have also developed our own code (in C, Python and GnuRadio) for analysing these data.

\section{Band integrated and long-term RFI characterisation}

Even though Murriyang is not designed as a WiFi and Bluetooth receiver, and can not point directly to the horizon (where the majority of these terrestrial signals originate), the RFI picked up in the side lobes do still impact observations. A typical Bluetooth transmitter outputs between 2.5\,-\,100\,mW \citep{Anees2024bluetooth}. If in the primary beam of the telescope (assuming 1\,s integrations), such a signal could be detected from a range of $\simeq$83,000--520,000\,km. Given the telescope's focus cabin height of 58\,m, the line of sight to the radio horizon is approximately 27.2\,km. This places the town of Parkes, located 17.8\,km away, well within range, and thus a potential source of RFI.

However, Murriyang has an elevation limit of 30.25 degrees and hence the signals from Parkes will be detected through the far side-lobes. In the far side-lobes with a gain of 0\,dBi, the range a Bluetooth signal could be detected is reduced to $\simeq$7,300--46,500\,km. 
The strongest individual signals are strongly correlated with the operating hours of the visitors’ centre (which typically receives up to around 100,000  visitors per year), from our own laboratories and on-site offices, and from the local farming community. 

\begin{figure}[hbt!]
\centering
\includegraphics[width=1.0\linewidth]{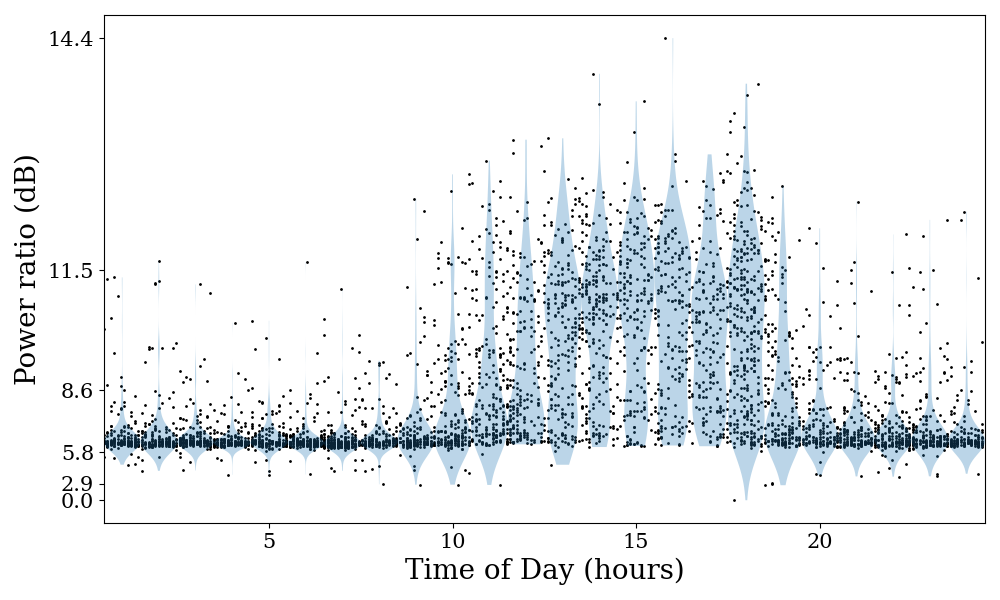}
\caption{Violin plot showing the total measured power in the 2432\,MHz subband (as a ratio of the minimum power recorded)  as a function of time of day, extracted from calibrator data files taken over the period of 5 years. It shows a distinct increase in the RFI around standard Australian business hours 9am to 5pm.}
%/DATA/CETUS_8/mar855/RFI_project/final_data_sets/python_plotters/power_ToD_plot.py
\label{fig_ToD}
\end{figure}

\begin{figure}[hbt!]
    \centering
    \begin{subfigure}[t]{1\linewidth}
        \centering
        \includegraphics[width=1.0\linewidth]{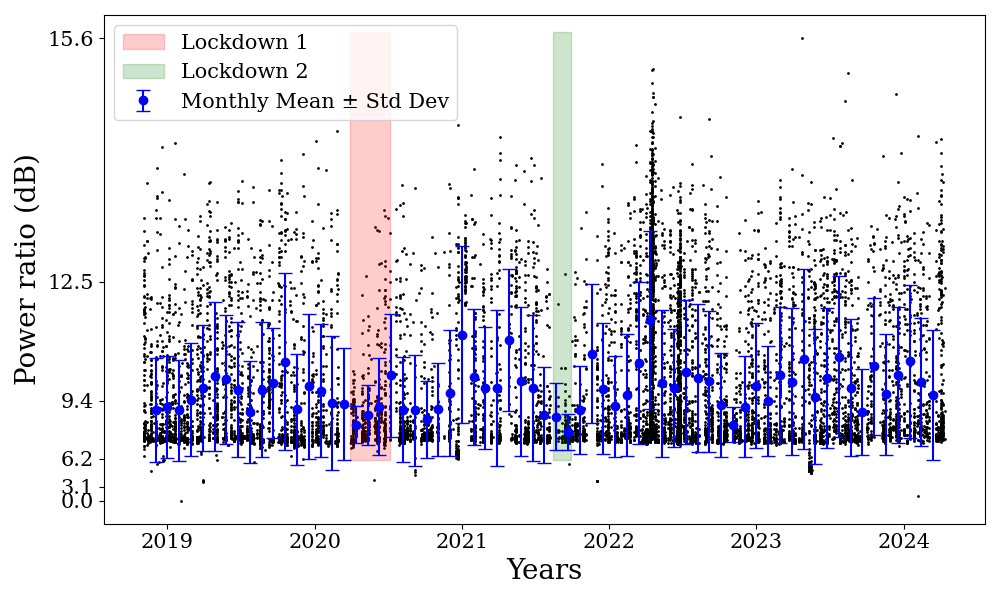}
    \end{subfigure}
    
    \begin{subfigure}[t]{1\linewidth}
        \centering
        \includegraphics[width=\linewidth]{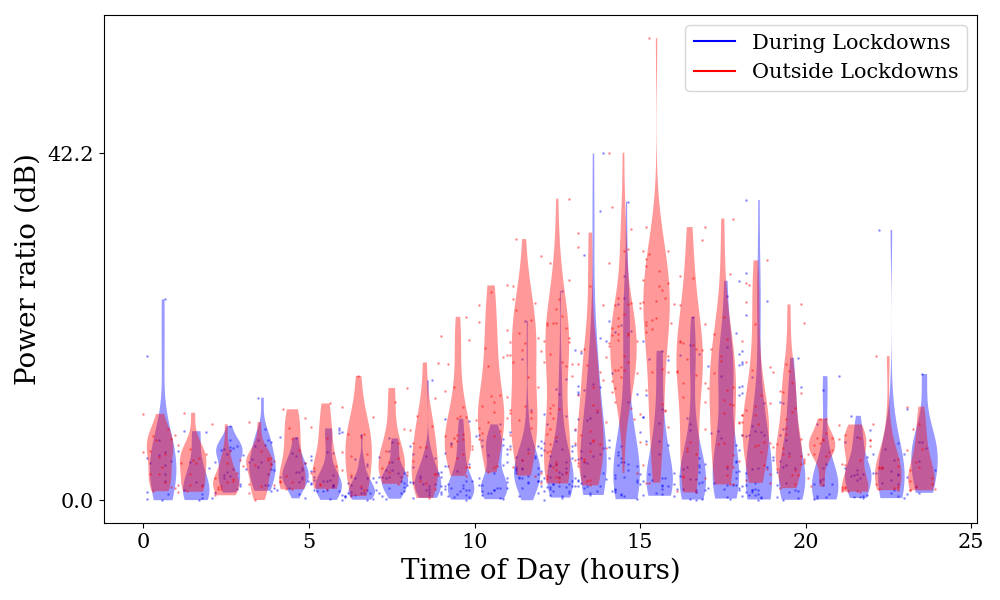} 
         
    \end{subfigure}
    
    \caption{The top panel shows signal strength received in the 2432\,MHz subband (as a ratio of the minimum power recorded) over a period of 5 years, over a restricted elevation angle of 40-60\textdegree. The Parkes Telescope visitor centre was closed due to lockdowns from the 27th March - 4th July 2020 (Lockdown 1) and 15 August - 27 September 2021 (Lockdown 2). The circles show the monthly mean received signal strength, as extracted from calibrator data files taken over the period of 5 years, with the bars showing $\pm$ 1 standard deviation. The coloured bars show the period of the lockdowns and a noticeable decrease in the intensity of the monthly mean of the RFI. 
    The lower panel shows the comparison in power as a function of time of day, this time comparing the power during the first lockdown (April to June 2020) and the power from an equivalent date range outside of the lock-down period (April to June 2023). For this panel we did not constrain the elevation angle to ensure an adequate number of observations for comparison.} 
    \label{fig_lockdowns}
\end{figure}

To investigate the origin of the 2432\,MHz subband RFI in greater detail, we analysed the archival calibrator observations by computing the average power across the subband 
as a function of time and pointing direction. As shown in Figure \ref{fig_ToD}, there is a clear increase in received power during typical Australian business hours (9am–5pm). There was also a noticeable decrease in the intensity of RFI during periods when the Parkes telescope visitor centre and on-site offices were closed due to local COVID lockdowns. These can be seen in Figure \ref{fig_lockdowns}, but even during such lockdowns some observations were affected by RFI suggesting, as expected, that not all the RFI originates from a single place.  The lower panel in Figure~\ref{fig_lockdowns} shows the decrease in power received during the first lockdown period. Note that this panel includes all data obtained during the specific time periods, whereas the top panel (and other related Figures) uses a restricted elevation angle from 40-60\textdegree.

\begin{figure}[hbt!]
\centering
\includegraphics[width=1\linewidth]{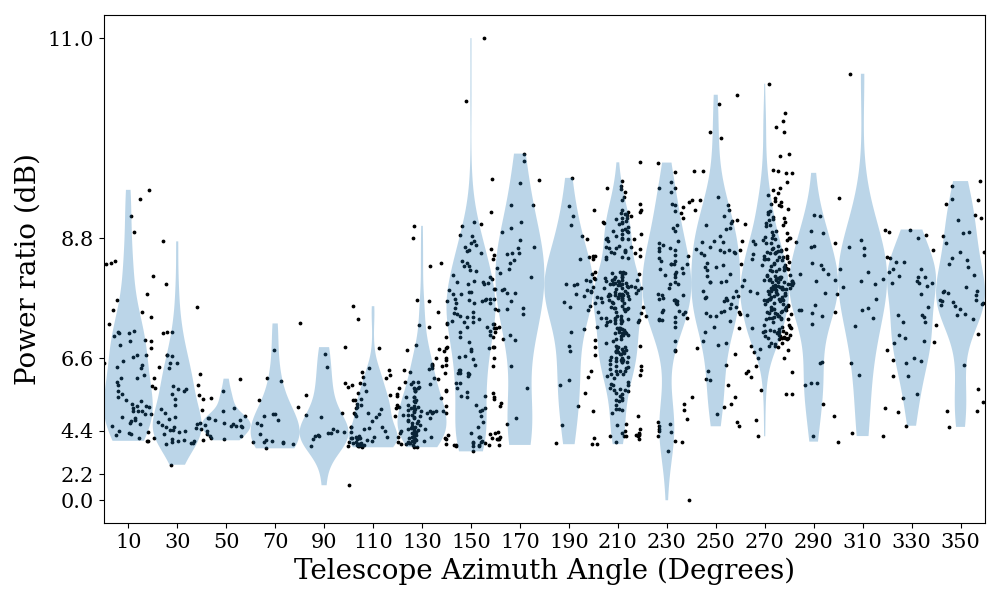}
\caption{Signal strength received by the telescope in the 2432\,MHz subband (as a ratio of the minimum power recorded) as a function of direction {\bf (in $20^\circ$ intervals)}, extracted from calibrator data files taken over the period of 5 years. This shows a significant, direction-dependent variation in power received, with significant increases to the West and relatively low levels in the range from 40--140\, degrees\,E. }
\label{fig_power_azimuth}
\end{figure}

\begin{figure}[hbt!]
\centering
\includegraphics[width=0.95\linewidth]{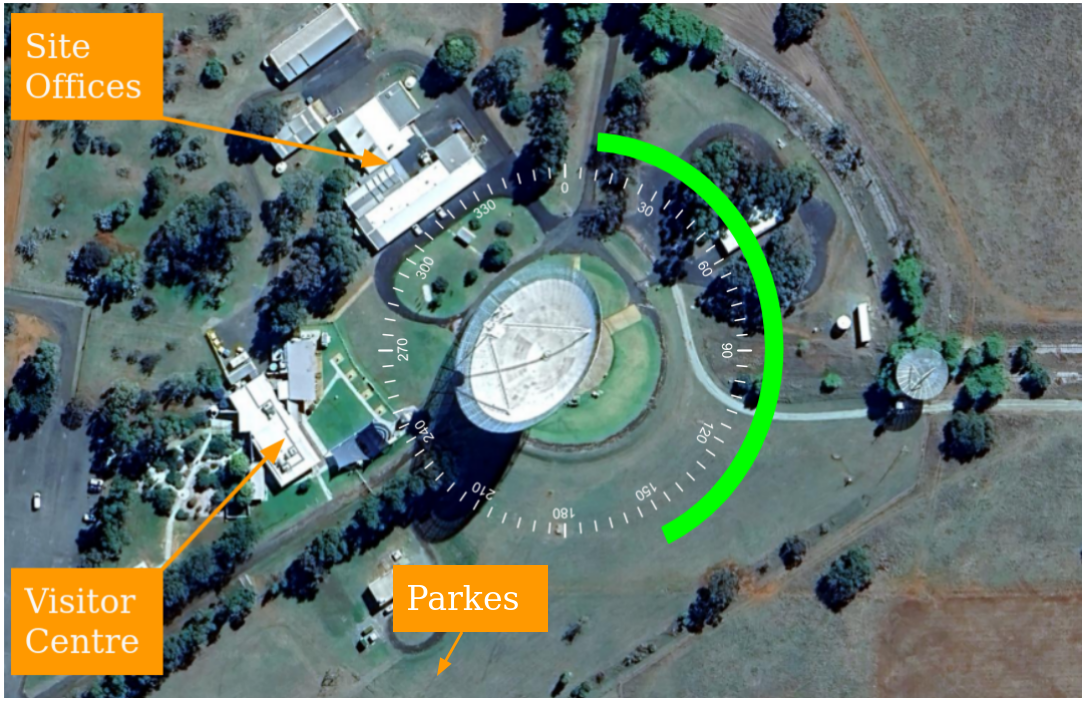}
\caption{Satellite image of the Parkes telescope site. The compass centred in the telescope shows that the location of the visitor centre (marked) is to the West of the telescope at W\,240\textdegree to W\,275\textdegree and the site office building is located between W\,290\textdegree to W\,340\textdegree. The green bar shows the range of angles from Figure \ref{fig_power_azimuth} with the lowest received signal strength. (Image: Google Earth)}

\label{fig_map}
\end{figure}

As shown in Figure \ref{fig_power_azimuth}, the RFI signal strength in the WiFi/Bluetooth bands shows a distinct directional bias to the West, which is in the direction of the town of Parkes, the Parkes Observatory Visitor Centre and Site Offices. Figure \ref{fig_map} shows a satellite image of the location of the various buildings and telescope on the observatry site. A noticeable drop in the power of the RFI received by the telescope can be seen from the angles E\,40\textdegree to E\,140\textdegree while conversely an increase in the power received can be seen between the angles of W\,150\textdegree to W\,360\textdegree. This is likely due to the 64 metre reflector screening the prime focus from the RFI source and provides additional evidence that the likely culprits for the sources of RFI are the visitors centre, site offices and/or Parkes. 

Even though visitors to the Parkes Observatory site are asked to switch off their Bluetooth and WiFi-enabled devices, it is clear that many do not do so. Visitors may also be setting their phones to Flight Mode in good faith, unaware that on newer mobile handsets this no longer automatically disables Bluetooth and Wifi. In some instances they may turn off a primary device (e.g. a phone) but not a secondary device (e.g. smart watch), which may result in the secondary device continuing to advertise and probe for connection. Noting that such devices are now ubiquitous, it is unrealistic to expect a pristine, RFI-free, site next to a visitor centre.  For observations which require clean 2.4\,GHz observations, night-time observations are less affected.  However, as we will show below, we can recover much of the astronomical signal in this band even during the day-time.

%%%%%%

\section{Temporal and frequency resolved RFI characterisation} \label{Section_characterising}

RFI caused by WiFi and Bluetooth should be characterised by their well-defined protocols.  In the 2.4 GHz band, the most common Wireless Local Area Network (WLAN) standard used today is the IEEE 802.11 family, specifically 802.11 and 802.11b/g/n/ax. This WiFi band runs from 2.401-2.4835\,GHz and is divided into 13 channels in Australia. Each channel is spaced 5\,MHz apart. The channel width depends on the signaling method, with Direct Sequence Spread Spectrum (DSSS) using 22\,MHz wide channels and Orthogonal Frequency-Division Multiplexing (OFDM) using 20\,MHz wide channels. This results in 11 overlapping channels, with only three non-overlapping channels usable at once, usually channels 1, 6, and 11.

The maximum length of a WiFi packet is 2304\,bytes plus encryption. Depending on the encryption protocol this could add 8--20\,bytes to this length resulting in maximum total package lengths of 2312--2324\,bytes. The bit rate for WiFi varies with the generation, with ``WiFi-0'' having a bit rate of 1--2\,Mbps and ``WiFi-7'' (the latest available protocol) having a theoretical bit rate of 46\,Gbps. This results in a possible packet length of 2324\,$\mu$s for ``WiFi-0'' and as short as 0.05052\,$\mu$s for ``WiFi-7''.

With the proliferation of Bluetooth connectable technology available today, the RFI is no longer limited to mobile phone handsets, but also includes wearable devices such as watches and fitness monitors, and also cars. Bluetooth transmits in one of two fashions: (1) the Basic Rate/Enhanced Data Rate (BR/EDR; also called Classic Bluetooth) which uses 79$\times$1\,MHz channels and (2) the Low Energy (LE; or BLE) which uses 40$\times$2\,MHz channels\footnote{https://www.bluetooth.com/learn-about-bluetooth/tech-overview/}. Both methods use frequency hopping to transmit data, switching channels pseudo-randomly 1600 times per second in a connected state and 3200 times per second in an inquiring state, resulting in packet lengths of 312-625 $\mu$s.  This frequency hopping reduces signal interference  with other Bluetooth signals in this highly used unlicensed portion of the ISM band, but can interfere with WiFi channels. Compared to WiFi, Bluetooth is short-ranged and used to establish Personal Area Networks (PAN) that are limited to 100m for Class 1 Bluetooth devices and 30-40m for Class 2 Bluetooth devices (generally available in commercial devices such as phones).

Bluetooth Low Energy advertising channels are used by devices to establish connections with other devices. There are three Primary Advertising Channels that are selected to minimise interference to the primary WiFi data channels. The three channels are Channel 37 centred on 2402\,MHz, Channel 38 centred on 2426\,MHz, and Channel 39 centred on 2480\,MHz. Advertising data is transmitted simultaneously on all three channels to increase the probability of receipt. Since all devices using Bluetooth Low Energy advertise for connection on the same three advertising channels, these channels usually contain a major part of the Bluetooth RFI. This also indicates the predominance of Bluetooth RFI appears to be coming from devices using the LE protocol.

\subsection{Demodulating the Bluetooth signal}

\begin{figure}[hbt!]
\centering
\includegraphics[width=1\linewidth]{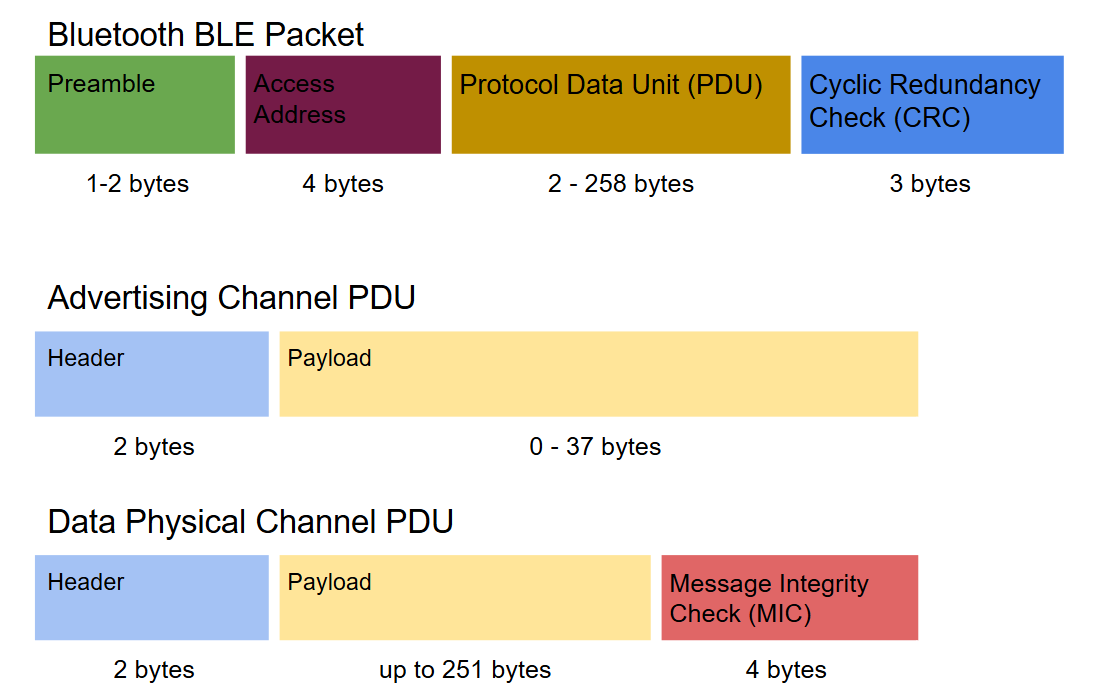}
\caption{The structure of different Bluetooth packets. At top is the whole packet structure, while the lower two parts show the different forms for the Protocol Data Unit (PDU) for the top figure.  }
\label{fig_BLE_packets}
\end{figure}

Bluetooth uses a Gaussian Frequency Shift (GFSK) modulation scheme to convert each channel's carrier frequency to transfer binary information. This is done by modulating the carrier frequency by at least $\pm 185$KHz with an increase representing a 1 (mark) and a decrease representing a 0 (space)\citep{Anees2024bluetooth}. Each Bluetooth packet is made up of four distinct parts: a preamble, access address, Protocol Data Unit (PDU) and a Cyclic Redundancy Check (CRC), see Figure \ref{fig_BLE_packets}. The PDU of the packet will differ based on whether the device is requesting a connection on an advertising channel or in a connected state and transferring data  (see Figure \ref{fig_BLE_packets}). Regardless, the preamble of a packet is a string of alternating 1s and 0s whose purpose is, among other things, to set the symbol timing estimation which essentially informs the receiver of the bit rate transmission. 

The Bluetooth signal packets seen in our recorded data ranged in time from around 288--25000 $\mu$s while the WiFi transmissions ranged in time from around 1200--3500 $\mu$s. Even though the Bluetooth protocol allows for transmissions between 312--625\,$\mu$s in length, our data shows that the majority of the Bluetooth packets lasts around 288\,$\mu s$.

\begin{figure}[hbt!]
\centering
\includegraphics[width=1\linewidth]{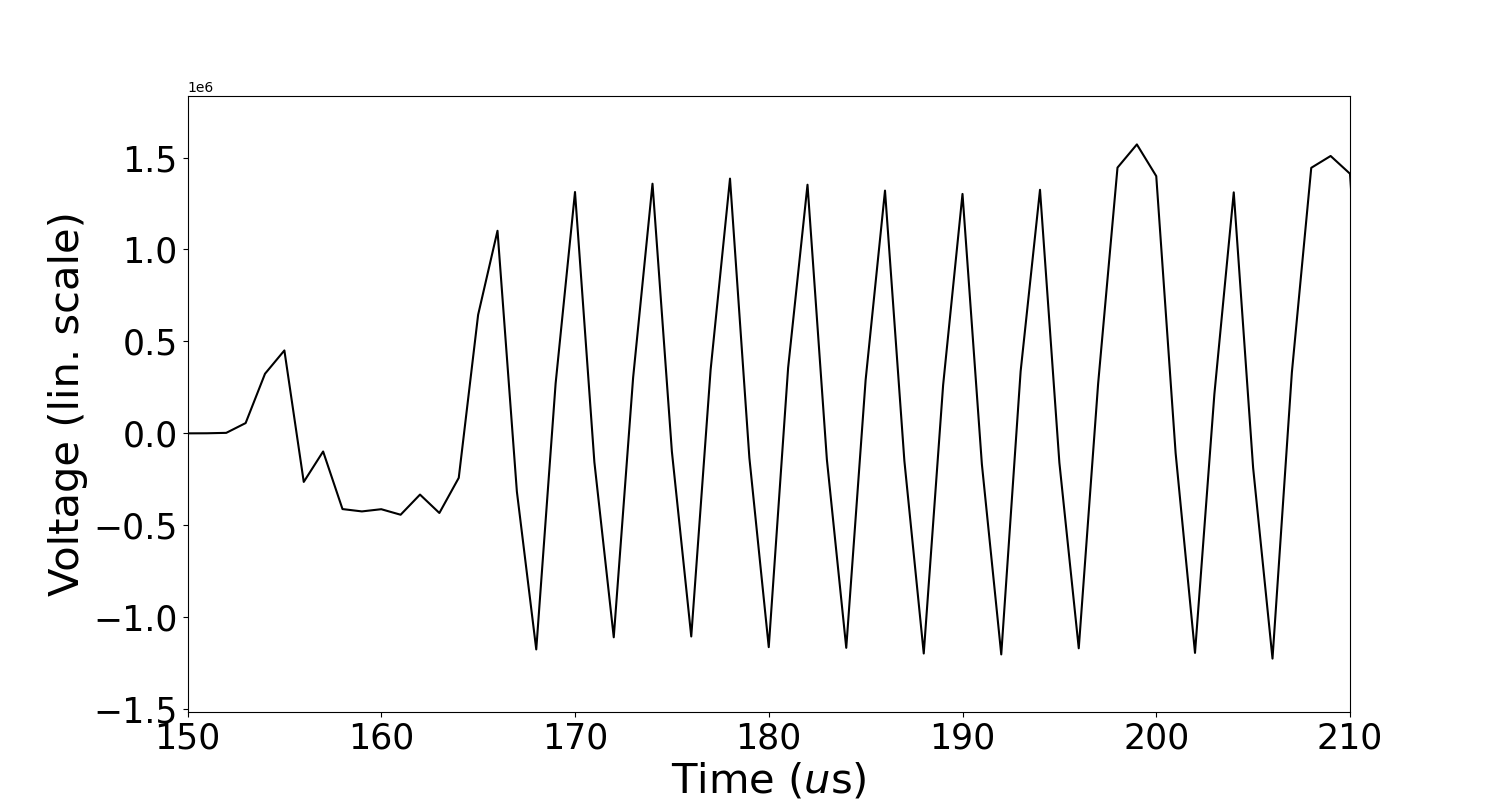}
\caption{The beginning of a representative Bluetooth advertising packet after the signal has been demodulated. This shows the 2402\,MHz advertising channel (Channel 37) from 150 $\mu$s in our data. This shows the bits encoded as Gaussian Frequency Shift Keying from the central frequency. The preamble can be seen by the alternating marks (1s) and spaces (0s). }
%image made using john_code_mine.ipynb on my laptop
\label{fig_BT_demod}
\end{figure}

We developed tools to filter the incoming data stream to select a specific Bluetooth channel, identify regions containing transmission and then to demodulate the packet, see \ref{app1}. The code, working on each individual BLE channel, shifts the Bluetooth frequency channel to baseband, applies a bandwidth Finite Impulse Response (FIR) filter to the signal, downsamples the voltage data to closer to the bit rate of the Bluetooth signals, and finally performs a simple demodulation to obtain bits of information from the carrier frequency. As an example Figure \ref{fig_BT_demod} shows the beginning of a representative demodulated packet with the triangular pattern indicating the 10101010 header information.  

\begin{figure}[hbt!]
\centering
\includegraphics[width=1\linewidth]{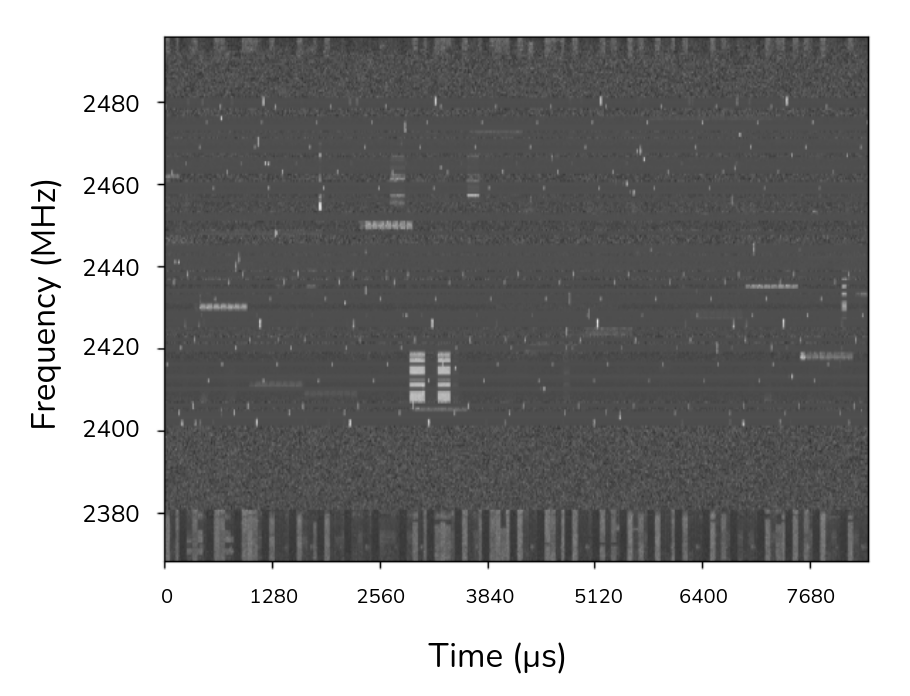}

\caption{Dynamic spectrum of a 0.5\,s sample of the channelised data from our observation. Visible in the data are samples of Bluetooth and Wifi. The narrow frequency band features (horizontal), both shorter and loner time spans, are Bluetooth signals. The intermediate time span signals over broader frequency ranges (vertical) ar WiFi signals. The RFI in the lower frequency range is due to NBN and is aliased at the upper edge of the frequency band.  }
\label{fig_time_freq}
\end{figure}

\begin{figure}[hbt!]
\centering
\includegraphics[width=1\linewidth]{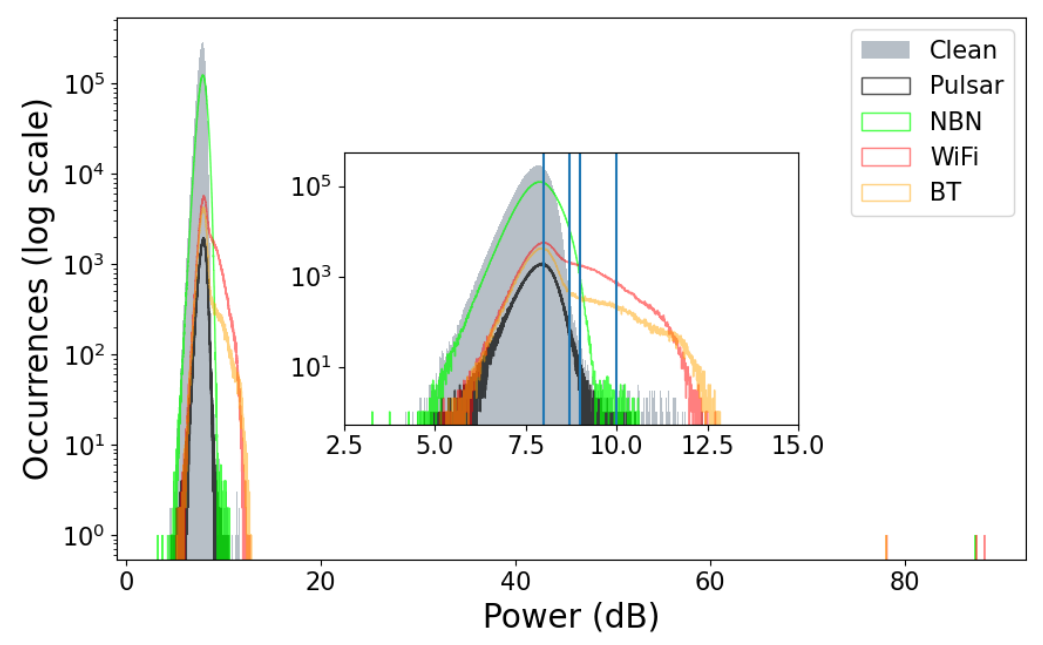}

\caption{Histogram showing the number of samples received as a function of power in the the 2432\,MHz subband in the channelised time series. Both the x and y axes are on a log scale. Colours show some of the elements in 1.3\,s of the channelised data including clean signal in solid grey, the pulsar signal in black, the NBN RFI in green, WiFi RFI in orange and Bluetooth RFI in yellow. As described in the text, all categories contain significant `contamination' from clean samples, but the characteristic power levels of the brightest RFI can still be seen. The blue lines represent the voltage thresholds that were tested at 8, 8.69, 9 and 10\,dB in Section \ref{sec: power threshold}.}
\label{fig_power_hist}
\end{figure}

\subsection{Characterising the components of the Parkes voltage data}

WiFi and Bluetooth signals that are significantly brighter than any reasonable expectation for an astronomical signal could potentially be mitigated using a simple  thresholding algorithm.  In order to explore the relative brightnesses of the RFI, the telescope system, and the bright astronomical source in our test data (the Vela pulsar), we channelised the 10\,s voltage data set with high time resolution (microsecond resolution), summed the polarisations and then inspected the resulting time-frequency output by eye.  Figure \ref{fig_time_freq} shows the distinct time and frequency signatures of the various RFI packets, the intermediate duration wider frequency RFI being WiFi packets, while the longer narrow frequency packets and the short duration narrow frequency packets are Bluetooth packets.   
The RFI covering the entire time period at the lower end of the frequency range is due to the NBN, and is also aliased in the upper end of the range. Within our 10\,s of data the majority of the RFI appears to come from the BLE advertising channels (but not all; we see Bluetooth transmission across the 2402--2480\,MHz band). 
Having demodulated each of the bluetooth advertising channels, we note that the device addresses being transmitted indicate that the signals are from independent devices searching for other Bluetooth enabled devices.

For a representative 1.3\,s of data, we manually selected regions that were clearly dominated by the various signals of interest (WiFi, Bluetooth, NBN transmissions, the pulsar and clean parts of the band) and measured the power levels of the samples in each category.  The results are shown in Figure \ref{fig_power_hist}. Note that this relatively-crude approach of manually selecting data blocks results in the inclusion of clean data in each RFI category, but is still informative. The RFI ranges in power from the level of the astronomical signal to orders of magnitudes higher.

We also note that the power levels are obtained from microsecond time resolution and hence include very short, very bright bursts of RFI\footnote{In our earlier analysis (Figures \ref{fig_ToD} to \ref{fig_power_azimuth}) we used datasets that have averaged a spectrum over minutes (typically $\approx$2 minutes) and hence the total power levels in those figures are significantly lower than the measurements from the high time resolution data.}.

Clearly the very brightest WiFi and Bluetooth signals could be removed with a simple thresholding algorithm with no loss of astronomical data. However, more care is needed to distinguish the weaker RFI from astronomical signals.  

\section{Mitigation} \label{section_mitigation}

In the standard astronomical data products available from the Parkes observatory (such as folding the data stream at the known period of a pulsar or calibrator, observing with high spectral resolution or with high time resolution), it is not currently possible to mitigate the Bluetooth and WiFi signals without simply flagging the majority of the band and hence losing the astronomical signal entirely in that band. 

The primary receiver currently installed on the Parkes telescope is a single pixel feed, and hence we cannot use mitigation methods such as beam nulling \citep{fridman2001rfi}.  We also do not have a fully implemented system to apply mitigation methods that rely on a reference antenna\citep{kesteven2005adaptive}.

\begin{figure}[hbt!]
\centering
\includegraphics[width=1\linewidth]{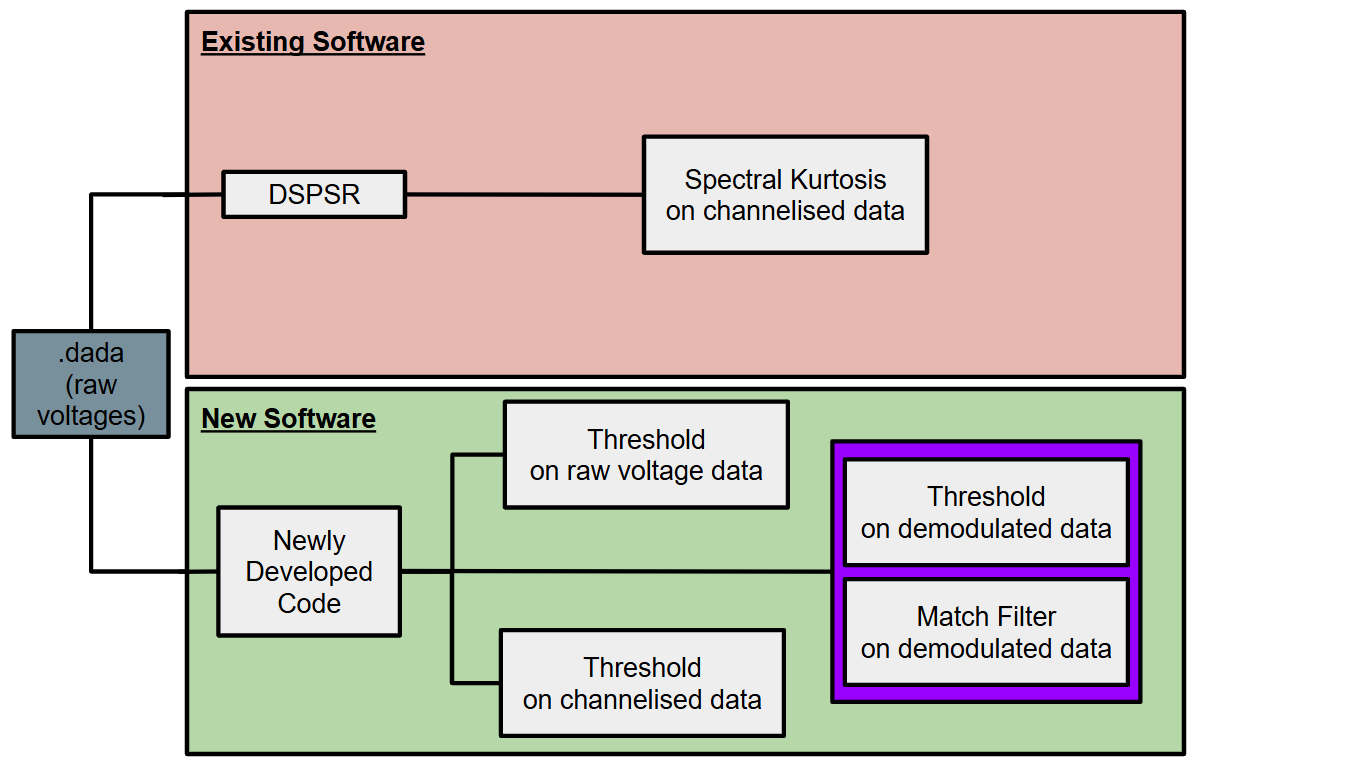}
\caption{Flowchart depicting general methods to approach the processing of the raw voltage data from Parkes observations. The Existing Software avenue includes the software package \textsc{DSPSR} and uses spectral kurtosis as the RFI mitigation method. Our new method uses software that mitigates RFI by applying a threshold to the channelised data, as well as applying filters to demodulated data streams.}
\label{fig_flowchart}
\end{figure}

We therefore have to rely on software methods for mitigation of the processed data products. There are two general avenues for this approach: (1) use existing software for RFI mitigation or (2) develop new software. The existing software package used for real-time processing of data from the Parkes Telescope is DSPSR. \textsc{DSPSR} uses spectral kurtosis for the mitigation of RFI in data that is folded at the period of a known pulsar or calibrator. Our newly developed software package  implements a number of other methods. Both DSPSR and the new package take voltage data streams as inputs, apply the mitigation algorithms and output astronomical data products.  The range of existing and newly-developed options tested in this work are represented in the flowchart shown in Figure~\ref{fig_flowchart}.

\begin{figure*}
    \centering
    \begin{subfigure}[t]{0.45\linewidth}
        \centering
        \includegraphics[width=\linewidth]{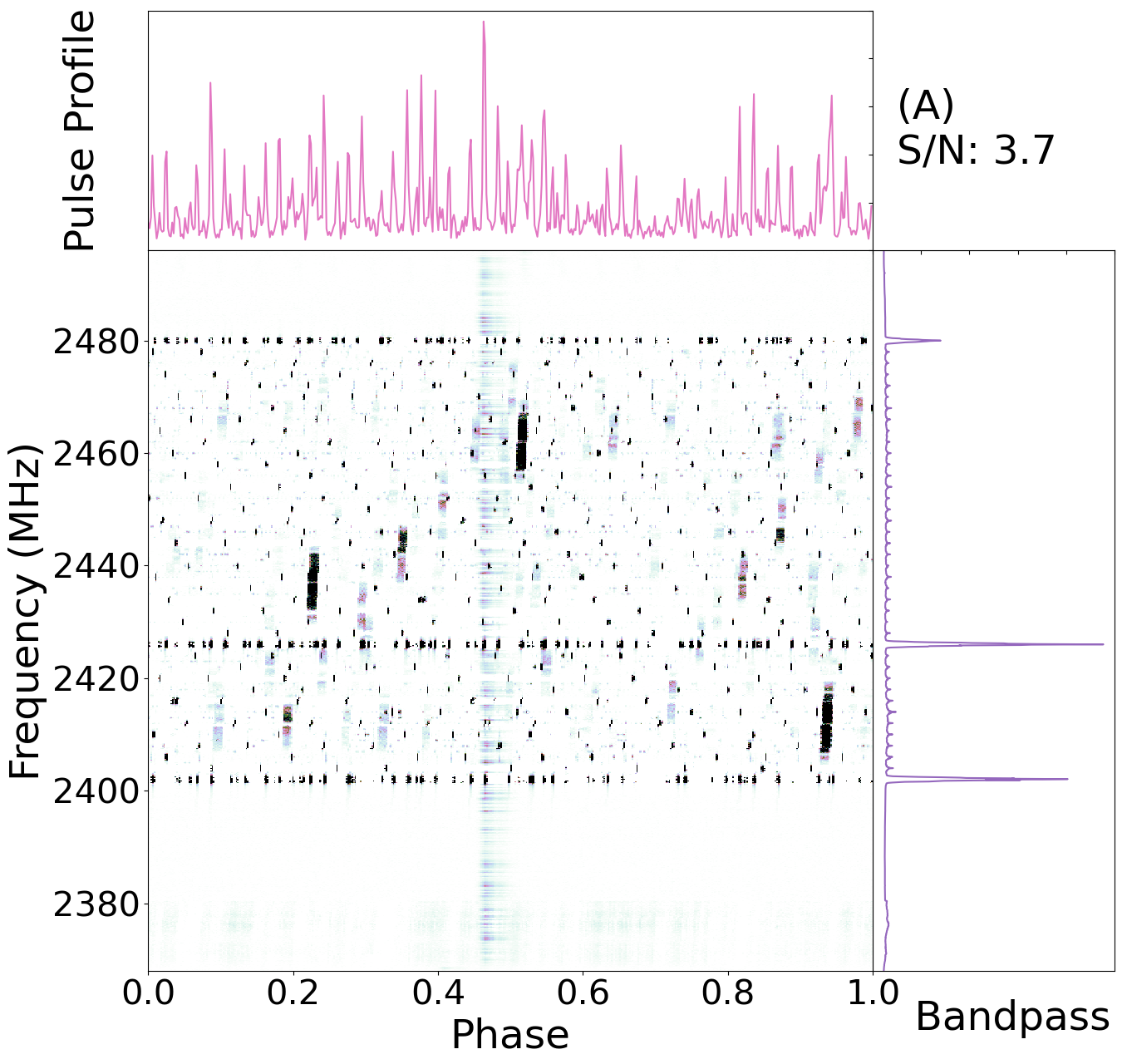}
    \end{subfigure}
    \hfill
    \begin{subfigure}[t]{0.45\linewidth}
        \centering
        \includegraphics[width=\linewidth]{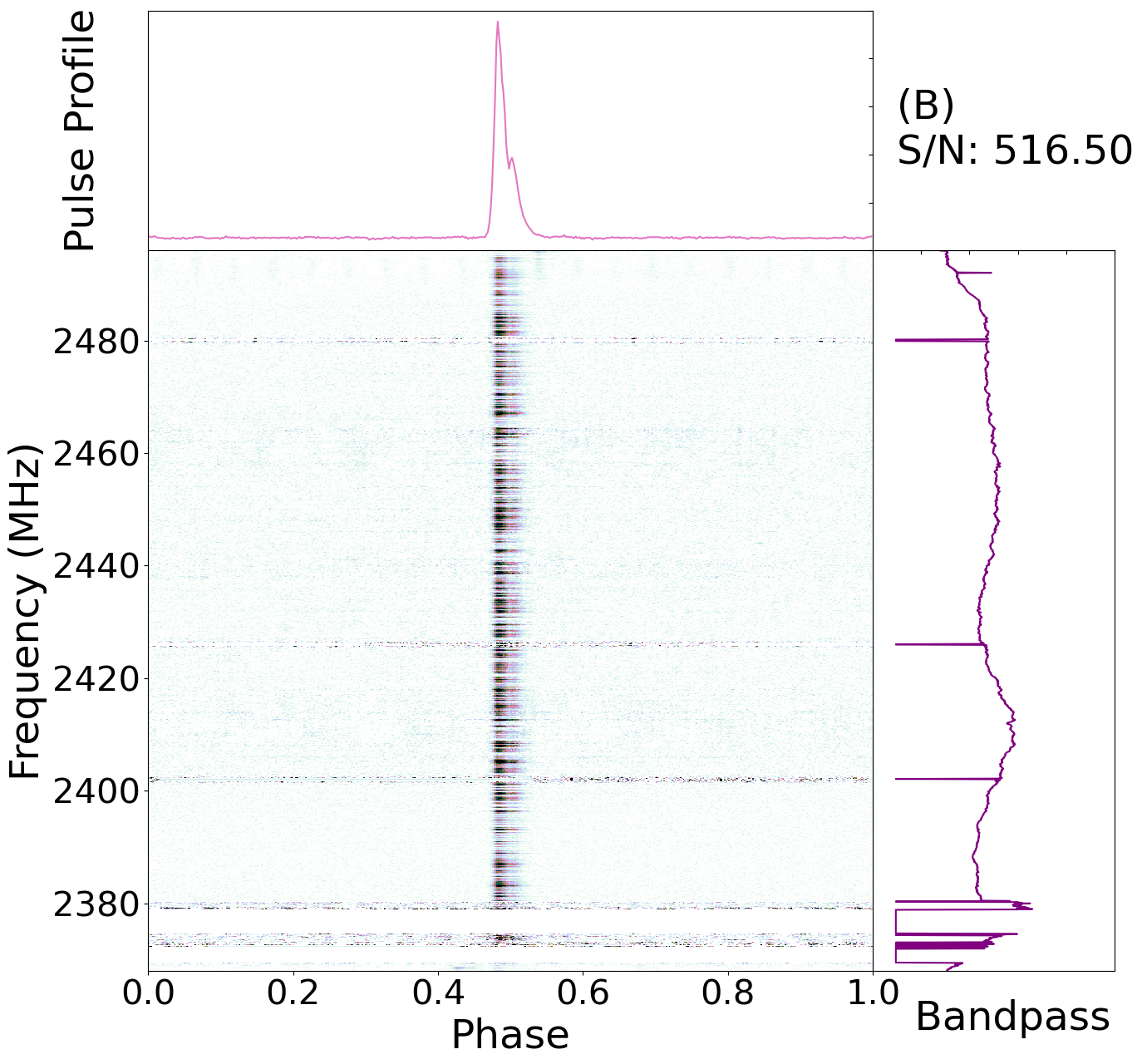} 
         
    \end{subfigure}
    \vspace{1cm}
    
    \begin{subfigure}[t]{0.45\linewidth}
        \centering
        \includegraphics[width=\linewidth]{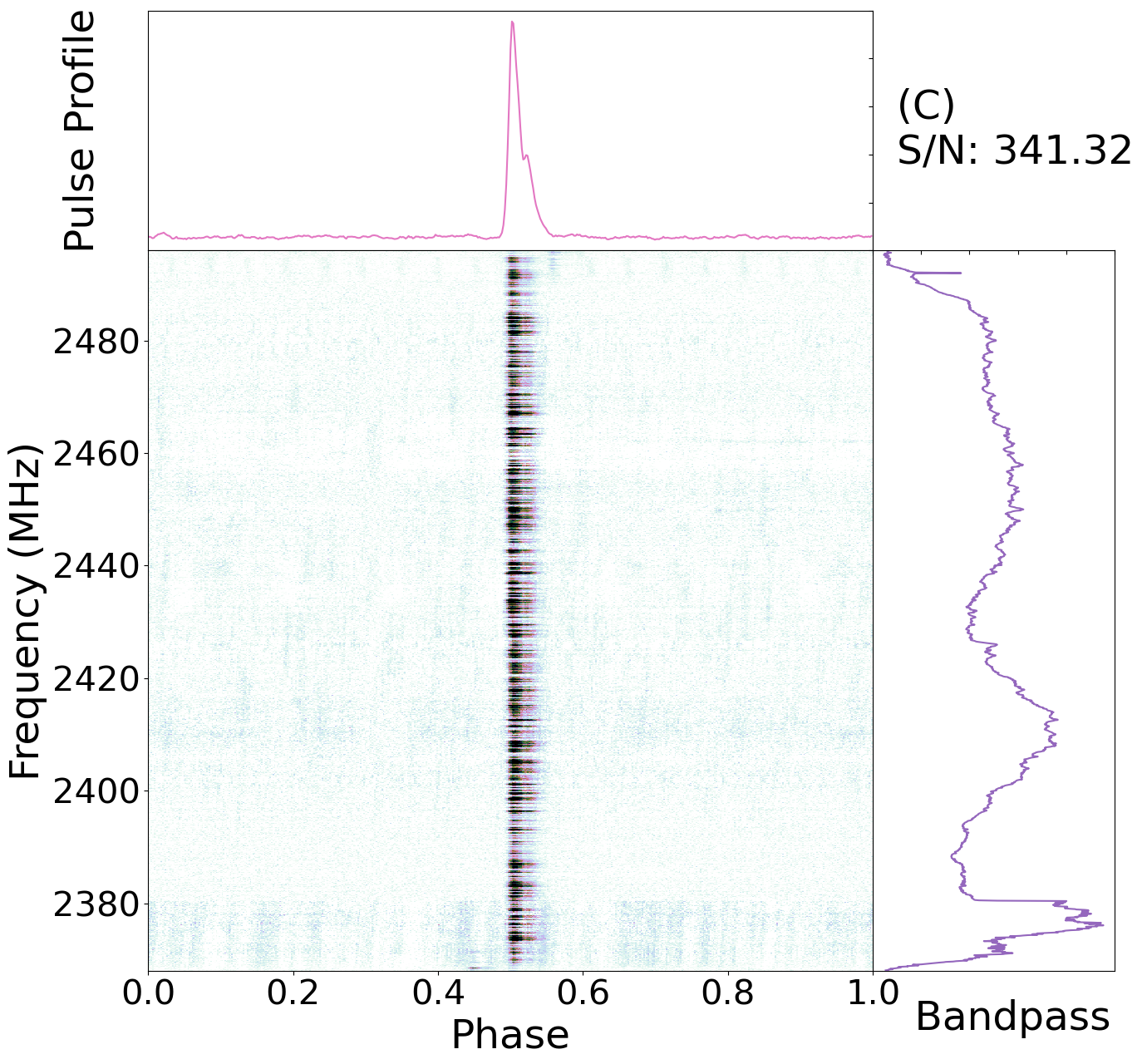}     \end{subfigure}
    \hfill
    \begin{subfigure}[t]{0.45\linewidth}
        \centering
        \includegraphics[width=\linewidth]{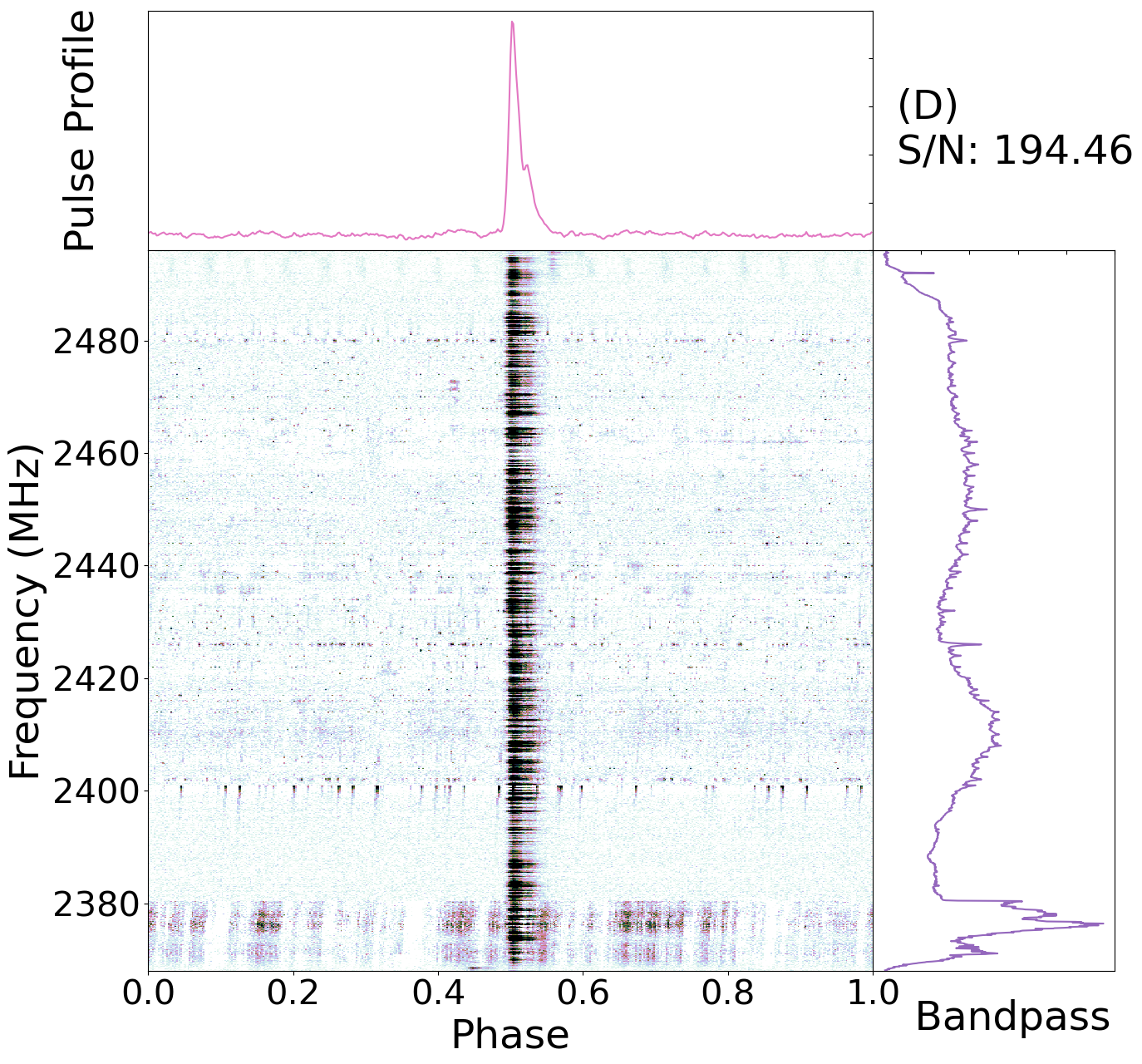} 
        
    \end{subfigure}
    \caption{The result of various mitigation methods. The power levels have been normalised.(A) The full 10\,s of channelised data with no RFI-mitigation applied. The dynamic spectrum is overpowered by the RFI. There is no discernible pulse profile and bandpass in dominated by powerful channelised RFI. (B) The folded dynamic spectra, pulse profile and bandpass with the default \textsc{DSPSR} spectral kurtosis settings ($N_{\sigma}$($\sigma$) = 4.1; $N$ = 2688; with no time or freq. scrunching) implemented. (C) The 10\,s of data after a 8.69\,dB power threshold has been applied. As a result there is a prominent pulse visible in the dynamic spectrum. There is also a visible pulse profile and relatively smooth bandpass present.  (D) The application of a combination of the averaged power threshold and the simple threshold on the demodulated data.}
    \label{fig:fourpanel}
\end{figure*}

Throughout the following, we compare the effectiveness of the various mitigation methods by folding the full 10\,s RFI-mitigated voltage data stream at the period of the Vela pulsar and measuring the resulting S/N of the pulse profile. We also display the folded pulse profile, phase-frequency diagram of the pulse profile, and the averaged (in time) frequency spectrum for all methods in Figure \ref{fig:fourpanel}.  For comparison purposes, Figure~\ref{fig:fourpanel}(A) shows the pulse profile without any mitigation (top panel). The pulse in this Figure is undetectable. The right-hand panel of the same figure shows the time-averaged bandpass with the three primary features being the Bluetooth advertising channels.  The central panel shows the de-dispersed pulse profile in a frequency-phase plot.  To indicate that the Vela pulsar is present we have had to choose a colour-scale in which the RFI is saturated.  The S/N is determined as the sum of the pulse profile (with prior knowledge of the phase of the pulse) divided by the standard deviation of the baseline. In this case the S/N value is dominated by the RFI.

\subsection{Spectral Kurtosis}

As a spectral kurtosis method is readily available via the DSPSR package, we demonstrate the effectiveness of this method first\footnote{The spectral kurtosis mitigation method is available to observers with the Murriyang, Parkes radio telescope. However, as of the time of writing there is a bug in the version of dspsr that means the output data product is not being produced and the variations in the RFI properties across the entire UWL observing band means that the parameterisation varies between bands.}.    Spectral kurtosis ($\widehat{SK}$) is the spectral variant of time-domain kurtosis \citep{dwyer1984use, vrabie2003spectral, nita2007radio, nita2010statistics}. Time-domain kurtosis measures the fourth statistical moment of a signal’s amplitude distribution to quantify its deviation from Gaussianity, and is commonly used for detecting transient acoustic events or diagnosing mechanical faults. In contrast, spectral kurtosis characterises the non-Gaussianity of signal power as a function of frequency, enabling the detection of impulsive or transient features in the spectrum. This property makes it a powerful and widely used tool for identifying radio frequency interference (RFI) in radio astronomy data.

\textsc{DSPSR} processes the raw voltage data by channelising it and then converting the signal to a detected power. The $\widehat{SK}$ estimator is computed from blocks of this detected power data to estimate its non-Gaussianity and determine  whether to excise it. Since most RFI exhibits  non-Gaussian statistics, with  power values skewed to higher levels, the $\widehat{SK}$ estimator can identify regions affected by interference \citep{nita2007radio}.  

The key input parameters for the \textsc{DSPSR} spectral kurtosis algorithm are the number of samples ($N$) used to compute the statistical estimates, and the threshold, expressed in units of standard deviations ($\sigma$), that determines whether a block of samples should be excised. 

\begin{figure*}
    \centering

    \begin{subfigure}[t]{0.45\linewidth}
        \centering
        \includegraphics[width=\linewidth]{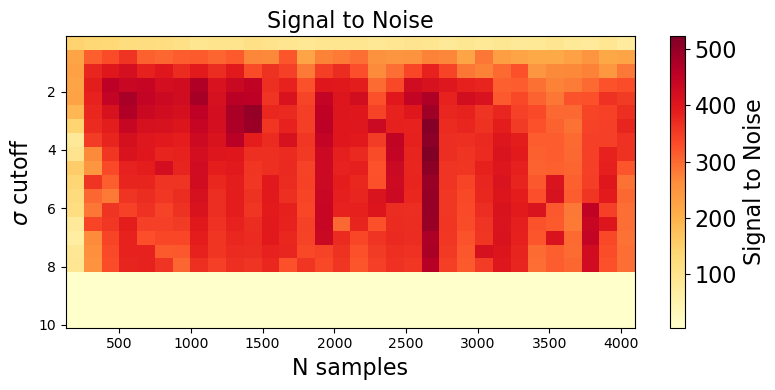} 
         
    \end{subfigure}
    \hfill
    \begin{subfigure}[t]{0.45\linewidth}
        \centering
           \includegraphics[width=\linewidth]{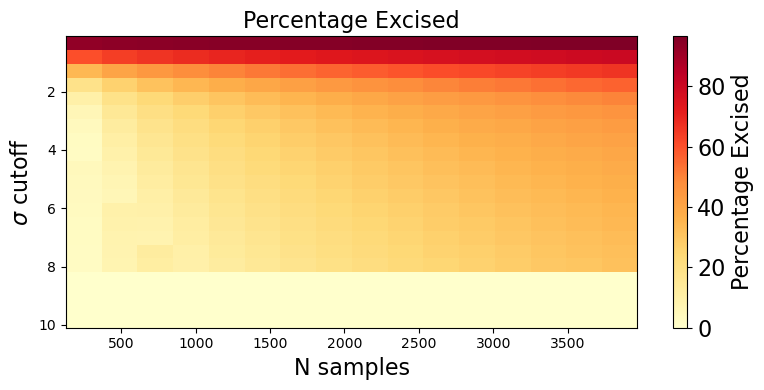}     
    \end{subfigure}
    \caption{Heat maps comparing the signal to noise ratios and percentage of data excised for the pulse profile output of \textsc{DSPSR}. Each frame shows the effect of varying the number of samples $N$ and the threshold for excision $\sigma$  used for calculating spectral kurtosis of RFI. These were calculated using 2048 frequency channels.}
    \label{fig:heatmaps}
\end{figure*}

We applied the kurtosis method with a range of trial parameters on our 10-s voltage data stream.  To compare the effectiveness of the algorithm we channelised the data into 2048 channels (note that this gives us a sample resolution of 62.5\,ms) and folded at the period of Vela with  512 bins.  We then measured the signal-to-noise of the resulting pulse profile and obtained the percentage of the 10\,s data that was removed. The results are shown in Figure~\ref{fig:heatmaps}\footnote{The DSPSR command used was: \\dspsr -no\_dyn -d 4 -F 2048:D -b 512 -skz -skz\_no\_fscr -skz\_no\_tscr -skzs <$\sigma$> -skzm <$N$> <filename.dada> }.  In all cases the S/N of the pulse profile significantly exceeds that obtained with no mitigation applied.  Comparing Figure \ref{fig:fourpanel}(A) with (B) shows that this mitigation method allows the Vela pulsar to be easily detected with high S/N. This Figure was obtained with 2048 frequency channels and processed with the parameters that give the maximum S/N (516) in Figure~\ref{fig:heatmaps}, obtained with a sample size of 2688 and a sigma threshold of 4.1. We note that this performance significantly exceeds that of the current DSPSR default settings ($N=128$; $\sigma=3$), with a S/N in these trials of 121.

\subsection{Simple Power Threshold}\label{sec: power threshold}

The simplest mitigation method is simply to inspect the 16-bit voltage data stream and skip blocks of data where the data values exceeds a given threshold.  However,  as the RFI signal hops in frequency we can significantly improve the algorithm by searching for strong signals in the known WiFi and Bluetooth bands after channelising and detecting the signal.  For a given frequency channel, signals above a specified power threshold are then not included when binning the pulse profile.

We  tested several power thresholds (10, 9, 8.69 and 8\,dB) defined as the ratio of the power to the minimum total power in the channelised data. The results are shown in Table~\ref{tab:results}.  The folded pulse profile obtained using the highest S/N set of parameters is shown in Figure~\ref{fig:fourpanel}(C).  We note that the resulting S/N of the profile is high and yet we only excise 0.4\% of the data (significantly less than the kurtosis method). The reduction in S/N of the pulse profile using this method compared to the spectral kurtosis is because low-level interference remains and hence the noise level of the baseline is larger in this scenario.

\subsection{Threshold on the Demodulated Bluetooth Signal}

Since each Bluetooth packet should begin with a series of eight alternating 1's and 0's (10101010 or 01010101), it should be possible to excise BLE packets by identifying this pattern in the demodulated signal and then removing a pre-determined sized packet of the subsequent data. For this approach, we demodulated each of the Bluetooth channels independently (our algorithm to do this is given in \ref{app1}). 

We then searched the output series for the sequence of eight 1's and 0's with a power threshold applied to prevent random occurrences of the sequence.  This technique led to 1.2\% of the data being excised, but the resulting pulse profile having a signal to noise ratio of 185.

This method leaves clear RFI in the data. This is because the method above only accounts for Bluetooth packets starting with the correct preamble. Other forms of RFI (including WiFi and NBN signals) may not necessarily have a preamble matching this pattern. We, therefore, also test applying a simple power threshold in the demodulated data. For this, we average across blocks of 64 samples of demodulated data then apply a power threshold to these blocks to eliminate some of the remaining RFI that was not removed by the application of the demodulated threshold. This combination of thresholds in the demodulated data resulted in 1.5\% of data being excised and a slightly improved signal to noise ratio of 194. Table \ref{tab:results} shows the comparison of percentage of data excised and signal to noise ratio of the Vela puslar between different methods of RFI mitigation.  The combined approach is shown in Figure \ref{fig:fourpanel}.

\begin{table}[ht]
\centering
 \begin{tabularx}{\textwidth}{l l l }
\hline
\textbf{Parameters} & \textbf{Percentage Excised} & \textbf{Relative S/N} \\
\hline
\multicolumn{3}{l}{\textbf{No Mitigation}} \\
  & 0\% & 3.7\\ 

 \multicolumn{3}{l}{\textbf{Spectral Kurtosis}} \\
 Default; $N$: 128 $\sigma$: 3 & 4.4\% & 121\\ 
 Highest S/N;  $N$: 2688 $\sigma$: 4.1 & 33.4\% & 516 \\ 
 \multicolumn{3}{l}{\textbf{Simple Power Threshold}} \\
 10\,dB & 0.04\% & 48.6 \\ 
9\,dB & 0.2\% & 212 \\ 
8.69\,dB & 0.4\% & 341 \\ 
8\,dB & 33.0\% & 304\\ 
\multicolumn{3}{l}{\textbf{Threshold on Demodulated Data}} \\
 Power threshold only& 1.2\% & 185 \\ 
 Power and Averaged threshold & 1.5\% & 194 \\ 
\hline
\end{tabularx}
\caption{This table compares the percentage of excised data and the relative signal to noise ratio for different mitigation methods tested.}
\label{tab:results}
\end{table}

\section{Discussion and Conclusion}
We see that for most of the methods, the outcomes are similar. By removing a few percent of the samples the astronomical signal (the Vela pulsar in this example) becomes useable, and the data can be used for e.g., pulsar timing experiments.

For pulsar fold-mode, continuum and spectral-line observations it is relatively straightforward to implement such a power threshold algorithm.  The rejected samples in the voltages would be identified and then not included in the folding or averaging process. For pulsar search-mode data the situation is more complex. The large data volume means that it is not common to provide a data weight or flagging information on a sample-by-sample basis. For search-mode observations quantised with a small number of bits, it would usually be preferable to replace the affected samples by representative noise, but the noise properties would need to be carefully chosen because of the fluctuating baseline.

We chose to demonstrate these algorithms through observations of the Vela pulsar.  Vela is so bright that even in 10\,s of data we can detect the folded pulse with high S/N. As the brightest pulsar (in this frequency band), any threshold that does not remove the Vela pulse is unlikely to remove or significantly affect any other pulsar. However, some other astronomical sources (such as FRBs and masers) can be much brighter.  Any online mitigation method based on thresholding should therefore allow those threshold values to be changed based on the science case, or for the RFI-mitigation to be switched off if the science case is to search for very bright events. (For such bright events the algorithm applied to the demodulated signal would be preferable as it searches for specific preambles of the Bluetooth packets and only applies to the specific Bluetooth frequency bands.)

As all the algorithms tested here produce similar output data quality, the choice for any online implementation would be based on which one is simplest to implement, but we note:
\begin{itemize}
\item All the algorithms require a channelisation step to account for the WiFi and Bluetooth signals switching between frequency channels.
\item A simple threshold would require that the user has the ability to set that threshold depending on the science case.
\item The spectral kurtosis algorithm is already implemented with the \textsc{DSPSR} software, but the settings required for its use change in different observing bands. The parameterisation given above for the 2.4\,GHz band are linked to the properties of the WiFi and Bluetooth packets and would not be the correct parameterisation for other bands.
\item Demodulating the Bluetooth signals would potentially allow identification of the devices causing the RFI. This would make the RFI easier to track down, but also may lead to privacy and security concerns.  We note that we do not propose to store the output of any demodulated data sets in e.g., our data archive and hence such information would not be publicly available.
\end{itemize}

Other mitigation methods are possible.  For instance a WiFi/Bluetooth detector could be installed in the Parkes Visitor Centre and connected to the main observing system.  Co-incident voltage detections in both systems could then be used to mitigate the RFI. Of course, the Observatory could also investigate new and creative strategies to encourage and educate visitors to switch off their transmitting devices, potentially with some form of real-time display of detected RFI in the Visitors Centre itself .

Our work has shown that astronomers using the Parkes UWL receiver system could carry out observations in the 2.4\,GHz part of the band, but our algorithms have not yet actually been implemented on the live observing system. Over the coming months we will endeavour to implement at least one of the new threshold methods and make that available to the user community.

\begin{acknowledgement}
The ultra-wideband receiver project was primarily funded through an Australian Research Council Linkage Infrastructure, Equipment and Facilities (LIEF) grant, with additional funding obtained from the Commonwealth Scientific and Industrial Research Organisation (CSIRO). We thank everyone who has provided advice, updated their software packages, or helped build, install and/or commission the system. 

The Parkes radio telescope is part of the Australia Telescope National Facility which is funded by the Australian Government for operation as a National Facility managed by CSIRO.
\end{acknowledgement}

%\endnote in some journals will behave like \footnote; and \printendnotes will not output anything. 
\printendnotes

\bibliography{references}

\onecolumn
\newpage
\appendix

\section{Bluetooth Demodulation Python Code}\label{app1}
\begin{verbatim}
#Bluetooth LE from UWL 16-bit baseband data

import numpy as np
import matplotlib.pyplot as plt
from scipy import signal
from scipy import io as signalIO
import os

#Read the file header
#The file header is 4096 bytes of ASCII describing the rest of the binary data in the file

filename = '<filename.dada>'
headerSize = 4096
printHeader = True
hdr = {}
fh = open(filename, 'rb')
headerBuf = fh.read(headerSize)
fh.close()
headerBuf = headerBuf.decode('utf-8')
for line in headerBuf.split('\n'):
    try:
        k, v = line.split(None, 1) # splits 
        each line in the header
        hdr[k] = v # first part of each line 
        in header is the key, other is the value for the dict
    except ValueError:
        pass

if(printHeader == True):
    print('HEADER INFO:')    
    for key, val in hdr.items():
        print(key, ':', val)

#Read in the raw data
#Data is 16-bits complex, 2 polarisations, offset binary encoding
#Alternating real/imaginary values
#Polarisations are in contiguous blocks of 2048 samples (2048 from polA then 2048 from pol B)

fs = 1.0/(float(hdr['TSAMP'])*1e-6) # sample rate in Hz
fc = int(hdr['FREQ'])*1e6 # centre frequency

file_stats = os.stat(filename)
totBytes = file_stats.st_size - 4096
sampBytes = (int(hdr['NBIT']) + int(hdr['NBIT']))/8
totDPSamples = totBytes/(int(hdr['NPOL'])*sampBytes)

print(f'File contains {totDPSamples:.0f} dual-pol samples ({totDPSamples/fs:.3f} seconds)\n')

nDpSamples = 65536
totBlocks = int(totDPSamples/nDpSamples)

BT_adv_chan = 2402e6 # Bluetooth adverisment channel 37
# frequency shift required to place the Bluetooth channel at DC in the sub-band
# the '-1' is needed because this sub-band is in the second Nyquist zone so
# the frequency axis runs backwards
fShift = -1*(fc - BT_adv_chan)
# time-domain modulation function to implement the frequency shift
modFunc = np.exp(1j*2*np.pi*fShift*np.arange(nDpSamples)/fs)

# load the FIR filter taps for the 1MHz BW low-pass filter (2MHz BW for DC-centred complex signal)
MATLAB_filter = signalIO.loadmat('lp_filt_1MHz_512Taps.mat')
ht = MATLAB_filter['ht'][0,:]

# construct the select index vectors to extract the alternating 2048-sample polarisation data blocks 
polAsel = np.zeros(nDpSamples,dtype=np.int_)
polBsel = np.zeros(nDpSamples,dtype=np.int_)
idxA = 0
idxB = 0
for lp in range(int(2*nDpSamples/2048)):
    if lp%2 == 0:
        polAsel[idxA*2048:(idxA+1)*2048] = range(lp*2048,(lp+1)*2048)
        idxA += 1
    else:
        polBsel[idxB*2048:(idxB+1)*2048] = range(lp*2048,(lp+1)*2048)
        idxB += 1

# open the .dada file and advance the file pointer past the first 4096 header bytes
fh = open(filename, 'rb')
headerBuf = fh.read(headerSize)

nBlocks = totBlocks
dataBufSize = int(int(hdr['NPOL'])*sampBytes*nDpSamples) # size in bytes (duap-pol samples, complex-values, 16-bits per value)
for i in range(nBlocks):
    dataBuf = fh.read(dataBufSize)
    data = np.frombuffer(dataBuf, dtype='uint16')
    # the next two lines convert the offset-binary format to Numpy floats
    data = data.astype(np.int32)
    data = (data - (2**15))
    dataCmplx = data[0::2] + 1j*data[1::2]
    polA = dataCmplx[polAsel]
    polB = dataCmplx[polBsel]
    
    if i == 864: # pick a block number that contains a BLE brst
        break
    
fh.close()

tm = np.arange(len(polB))/fs
plt.plot(tm*1e6, np.real(polB),label='Real component')
plt.plot(tm*1e6, np.imag(polB),label='Imaginary component')
plt.xlabel('Time ($u$s)')
plt.ylabel('Voltage (lin. scale)')
plt.legend()
plt.grid()
hFig = plt.gcf()
hFig.set_size_inches([15,8])
plt.show()

#Frequency shift, low-pass filter and down-sample the time-series then apply a phase demodulation

tsFreqShift = polB*modFunc
tsLPF = signal.lfilter(ht,1,tsFreqShift)
tsDS = tsLPF[::64]
tsDemod = tsDS[:-1] * np.conj(tsDS[1:])

plt.plot(np.imag(tsDemod),'.-')
plt.grid()
plt.xlabel('Time ($u$s)')
plt.ylabel('Voltage (lin. scale)')
hFig = plt.gcf()
hFig.set_size_inches([15,8])
plt.show()

#Decode the result

tmp = np.imag(tsDemod[166:710:2])
bitCnt = 0
bitSeq = ''
hexArray = []
for val in tmp:
    if val > 0:
        bitSeq = bitSeq + '1'
    else:
        bitSeq = bitSeq + '0'
    bitCnt += 1
    if bitCnt == 8:
        hexChar = f'{int(bitSeq,2):X}'
        hexArray.append(hexChar)
        bitCnt = 0
        bitSeq = ''

for i in hexArray:
    print(f'{i} ', end='')
\end{verbatim}
\end{document}